\documentclass[12pt,preprint]{aastex}
\usepackage{graphicx}

\newcommand{\elem}[1]{{\scriptsize~{#1}}}
\newcommand{\elemath}[2]{\textrm{{#1}{\scriptsize~{#2}}}}

\newcommand{\elesm}[2]{\textrm{{\scriptsize {#1}}{\tiny{#2}}}}
\newcommand{\phot}{~photons~cm$^{-2}$~s$^{-1}$~sr$^{-1}$}

\shorttitle{Exspected X-ray Emission from the Warm-Hot Intergalactic Medium}
\shortauthors{Ursino, E., Galeazzi, M., and Roncarelli, M.}

\begin{document}
\title{Effect of the Metallicity on the X-ray Emission from the Warm-Hot Intergalactic Medium}


\author{Ursino, E.\altaffilmark{1}, Galeazzi, M.\altaffilmark{1},  
	and Roncarelli, M.\altaffilmark{2}}
\affil{Physics Department, University of Miami, Coral Gables, FL 33155}
\affil{Dipartimento di Astronomia, Universit\`a di Bologna, via Ranzani 1, I-40127 Bologna, Italy}
\altaffiltext{1}{corresponding author, galeazzi@physics.miami.edu}


\begin{abstract}

Hydrodynamic simulations predict that a significant fraction of the gas in
the current Universe is in the form of high temperature, highly ionized plasma 
emitting and absorbing primarily in the soft X-ray and UV bands, dubbed the 
Warm-Hot Intergalactic Medium (WHIM). Its signature 
should be observable in red-shifted emission and absorption lines 
from highly ionized elements. To determine the expected WHIM emission in the 
soft X-ray band we used the output of a large scale hydrodynamic SPH simulation 
to generate images and spectra with angular resolution of 14'' and energy 
resolution of 1 eV. 

The current biggest limit of any hydrodynamic simulation in predicting the 
X-ray emission comes from metal diffusion. In our investigation, by using 
four different models for the WHIM metallicity we have found a strong dependence 
of the emission on the model used, with differences up to almost 
an order of magnitude. For each model we have investigated the redshift 
distribution and angular scale of the emission, confirming that most photons 
come from redshift $z<1.2$ and that the emission has a typical angular scale of 
less than a few arcminutes.

We also compared our simulations with the few currently available observations 
and found that, within the variation of the metallicity models, our predictions 
are in good agreement with current constraints on the WHIM emission, and at this time the
weak experimental constraints on the WHIM emission are not sufficient to exclude
any of the models used.

\end{abstract}


\keywords{methods: numerical, intergalactic medium, diffuse radiation, 
large-scale structure of universe, X-rays: diffuse background, radiation mechanisms: thermal}

\section{Introduction}
\label{introduction}

The problem of the missing baryonic matter in the nearby Universe 
is still open. On one side we have  
the observations of the Lyman alpha forest at red-shift $z=2$, the 
results from the WMAP experiment, and the prediction of the standard 
nucleosynthesis model. They all suggest a baryonic density $\Omega_B$, 
expressed as fraction of the critical density, 
being about $ 0.045$ \cite{Rauch98, Weinberg97, BurTyt98, Kirkman03, 
Bennett03, Spergel07, Komatsu09}. 
On the other side, from the observed mass distribution function of 
stars, galaxies and clusters, the measured baryonic fraction of the 
local universe comes out about 2 to 4 times lower \cite{Fukugita98}.
In recent years large-scale cosmological hydrodynamic simulations 
\cite{CenOst99, Croft01, Dave01, Yoshikawa03, Borgani04, CenOst06} allow 
us to trace the history of baryons from very high redshifts to $z=0$, and 
they all predict that almost $50~\%$ of the baryonic mass in the 
nearby Universe is in the form of a filamentary structure of gas at a temperature 
in the range of $10^5$-$10^7$~K, the Warm-Hot Intergalactic Medium (WHIM). 
This gas is highly ionized and can be observed in the soft X-ray band 
($0.1-1$~keV) and in the Far UltraViolet (FUV) band. Soft X-rays probe 
mostly gas at T$>10^6$~K, while FUV are a better probe for gas at 
T$<10^6$~K. If we assume thermal and ionization equilibrium, thermal 
emission models that comprise bremsstrahlung and emission lines 
\cite{Raymond77}, in the approximation of thin gas, predict that most 
of the emission from the WHIM (considering an average metallicity of 
the order of 0.1~Z$_\odot$) should be in the form of emission lines 
of highly ionized metals (O, N, C, ...).

Surveys of the sky in the soft X-ray band \cite{Snowden97, Mush00, 
McCammon02, Galeazzi07, Henley08} identify different contributions 
to the Diffuse X-ray Background (DXB), with local components like 
Solar Wind Charge Exchange, Local Bubble, and Galactic Halo, and 
extragalactic components, mainly AGN. A detailed inventory of the 
contribution from the different sources shows that there is still a 
fraction between 10\% and 20\% that could be due to a different source, 
such as the WHIM. 

Observations in both X-rays and FUV found evidence of O\elem{VI}, 
O\elem{VII}, and O\elem{VIII} absorbers from the Local Group WHIM 
\cite{Fang03, Nicastro02, Nicastro03, Rasmussen02, Sembach03, 
Williams05, Williams06, Williams07, Bregman09}. 
Observations with the \emph{Far Ultraviolet Spectroscopic 
Explorer}  and the \emph{Hubble Space Telescope} found almost a 
hundred O\elem{VI} absorbers at redshift $z<0.5$ along the lines of 
sight of $\sim30$ AGN \cite{Danforth05, Danforth08, Tripp08}, with a 
$dN_{\elesm{O}{VI}}/dZ\sim15$, together with 
N\elem{V}, C\elem{III}, C\elem{IV}, Si\elem{III}, Si\elem{IV}, and 
Fe\elem{III}. These absorbers can be associated to shock-heated gas 
at $T\sim10^{5.5}$~K, although there are claims that part of them 
could be associated to photoionized gas at $T<10^{4.5}$~K 
\cite{Tripp08}. Using 
O\elem{VI} as best tracer for this gas and assuming collisional 
equilibrium (CIE), ionization fraction $f(\elemath{O}{VI})=0.22$, and 
metallicity $0.1Z_\odot$ one obtains a detected $\Omega_{OVI\_WHIM}/
\Omega_B\sim0.1$ but, depending on the assumptions on metallicity, 
this fraction could be higher \cite{Danforth08}. This leaves up to 
$\sim50\%$ of low redshift baryons that still need to be accounted for 
\cite{Danforth08}. They are predicted to be in the form of the warmer 
WHIM ($10^6<T<10^7$~K), visible in the soft X-rays and having as 
primary tracers O\elem{VII} and O\elem{VIII}. 

X-ray observations are more difficult than in FUV 
because of instrumental limits. So far there are claims for only 
four detections in absorption. Two O\elem{VII} absorbers along the 
line of sight of Mrk 421 where identified with \emph{Chandra} 
\cite{Nicastro05}, but not confirmed by \emph{XMM-Newton} 
\cite{Rasmussen07}. 
An O\elem{VIII} absorber has been detected at $z\sim0.55$ towards PKS 
2155-304 \cite{Fang02, Fang07}, while another O\elem{VII} absorber has 
been identified in the Sculptor Wall at a redshift range of $0.028-0.032$ 
\cite{Buote09}. 

Detecting the WHIM in emission is even harder, since its contribution 
is at most $20\%$ of the total DXB and present time instruments are 
not suited to identify its weak signal. However it is possible to 
analyze global statistical properties of the WHIM, or to identify 
filaments of WHIM where they are expected to be very strong. Recent 
observations found the WHIM signature in the X-ray angular correlation 
function obtained from XMM-Newton images \cite{Galeazzi09}. Furthermore, 
an XMM-Newton pointing in the region between clusters Abell 222 and Abell 
223 found evidence of excess emission attributed to a filament connecting 
them \cite{Werner08}.

In this paper we use the cosmological hydrodynamic simulation by 
Borgani et al. (2004) to predict the X-ray emission from the 
WHIM. Compared to our previous work \cite{Ursino06}, 
we exploited the good spatial resolution (on the order of the adopted 
gravitational softening of $7.5$~h$^{-1}$~kpc, compared to the 
$\sim195$~h$^{-1}$~kpc resolution element used by Cen \& Ostriker 1999) 
of a Lagrangian simulation to create high resolution maps of the baryon 
distribution. 

We focused our attention on the effect of the metallicity model used
on the X-ray emission. Metallicity in the filaments is one of the greatest 
uncertainties in this type of simulations. Observations of the 
intergalactic medium put some limits on metals in the InterGalactic 
Medium (IGM). At $z\lesssim0.5$ groups and clusters have metallicity 
$\textrm{Z}\sim 0.3$~Z$_\odot$, while Ly-$\alpha$ clouds have 
$\textrm{Z}\sim0.1$~Z$_\odot$ with a large scatter 
($0.01<\textrm{Z/Z}_\odot<1$). Going at higher redshifts metallicity 
becomes lower, possibly by a factor 10 already at $z\approx3$ 
\cite{Finoguenov03, Prochaska04, Simcoe06}. From the analysis of the low redshift 
O\elem{VI} absorbers, colder WHIM structures have 
$\textrm{Z}\sim0.15$~Z$_\odot$ \cite{Danforth08}. Although there 
is a possible correlation between higher metallicities and higher densities, 
a simple modeling of IGM metallicity is difficult. At present time it 
is even uncertain if there was an early metal enrichment ($z\approx4$), 
if metals in the IGM are due to the newborn ($z<2$) galaxies, or if 
metal enrichment history is described by a mixture of the two models 
\cite{Aguirre05}. Cosmological simulations with an accurate physical description
(galactic winds, star formation, black hole feedback, and so forth) can 
come to an avail in reproducing the proper metallicity \cite{CenOst99b, 
Borgani04, CenOst06, Tornatore09, Wiersma09b}. These models predict
that the average WHIM metallicity 
is of the order of $0.1$~Z$_\odot$ but the behavior as a function 
of density changes widely from simulation to simulation. Due to the 
lack of tight constraints on metallicity from both observations and 
simulations, we allowed our code to work with a set of metallicity models, 
in addition to the one coming directly from the Borgani simulation. 
Thanks to this degree of freedom we were able to investigate the dependence 
of the soft X-ray emission on the different metallicity models.

The paper is structured as follows. In \S~\ref{Model} we introduce the 
hydrodynamic cosmological simulation, in \S~\ref{Simulation} we describe 
the code we used to simulate the X-ray emission in a selected field of 
view, in \S~\ref{Results} we show the simulated images and spectra, in 
\S~\ref{Metallicity-emission} we discuss how the emitted spectra depend 
on the metallicity models.

\section{The hydrodynamic Model}
\label{Model}

The general properties of the cosmological model are described in 
details in Borgani et al. (2004) and references therein. The 
simulation uses a flat $\Lambda$ cold dark matter ($\Lambda$CDM) model 
with cosmological constant $\Omega_\Lambda= 0.7$, 
$\Omega_\textrm{m} = 0.3$, and a baryon density 
$\Omega_\textrm{b} = 0.04$, the Hubble constant is $H_0=100$~h~km~s$^{-1}$
~Mpc$^{-1}$, with $\textrm{h} = 0.7$, and $\sigma_8 = 0.8$. The 
Plummer-equivalent gravitational softening was set as $\epsilon_{Pl}=
7.5$~h$^{-1}$~kpc. The code used to perform the simulation is the 
TREESPH code GADGET-2 \cite{Springel01, Springel05}. The simulation 
follows the evolution of $480^3$ dark matter (DM) particles and as 
many baryonic gas particles from redshift $z = 49$ to $z=0$. The box 
for the simulation is a cube of side 192~h$^{-1}$~Mpc, the DM and gas 
particles have initial masses $\textrm{m}_{\textrm{\tiny DM}} = 4.62 
\times 10^9$~h$^{-1}$M$_\odot$ and $\textrm{m}_{\textrm{\tiny gas}} = 
6.93 \times 10^8$~h$^{-1}$M$_\odot$ respectively. The physical processes 
involved in the simulation are gravity, non-radiative hydrodynamics, 
star formation, feedback from SNe with the effect of weak galactic 
outflows, radiative gas cooling and heating by a uniform, time-dependent, 
photoionizing ultraviolet background. 

The treatment of radiative cooling assumes an optically thin gas in CIE 
and uses only the primordial abundances (hydrogen mass fraction X$=0.76$, 
helium Y$=0.24$). Metals generated by the simulation itself are not 
considered for radiative cooling. A uniform, time-dependent UV background 
\cite{Haardt96} reionizes the Universe at $z\sim6$. 

Star formation is introduced following a hybrid multiphase model 
for the interstellar medium \cite{Springel03}. The interstellar 
medium, where star formation takes place, is represented as cold 
clouds (cold gas) embedded in a hot gas. Clouds are not accounted for individually 
in a given star-forming particle, but rather they are treated all 
together as a fraction of the total mass of the given star-forming 
particle. Every gas particle is considered as composed of two parts, 
the hot gas, with its own mass and density, and the cold cloud fraction, 
temperature and density determine the relative abundances of the 
two components. Whenever star formation takes place, a new star 
particle is spawned (with just a fraction of the starting gas 
particle), thus increasing the number of star particles. Stars are 
created, following a Salpeter initial mass function \cite{Salpeter55}, 
and instantly produce metals and release energy as supernovae.
Metals and energy are carried to the intracluster and intergalactic 
medium by galactic winds. Winds are introduced in the simulation as mass outflows 
with rate equal to twice the star formation rate and with a wind 
velocity of $360$~km~s$^{-1}$.

The output of the simulation consists of 102 boxes, equally spaced 
in the logarithm of the expansion factor between $z=9$ and $z=0$. 
As shown in previous work(e.g., \cite{Ursino06}) the X-ray emission 
above redshift 2 is negligible.  For our work we therefore only 
used the simulationup to redshift 2.

Besides the improvement in spatial resolution compared to our previous 
work, this simulation suites well our needs due to its large scale 
that allows us enough statistics for distant regions. More recent 
cosmological simulations have box sizes of at most 100~h$^{-1}$~Mpc 
\cite{CenOst06, Oppenheimer08, Tornatore09, Wiersma09b}, 
reducing the simulated sky area by at least a factor 3.

\begin{figure}
\begin{center}
\includegraphics{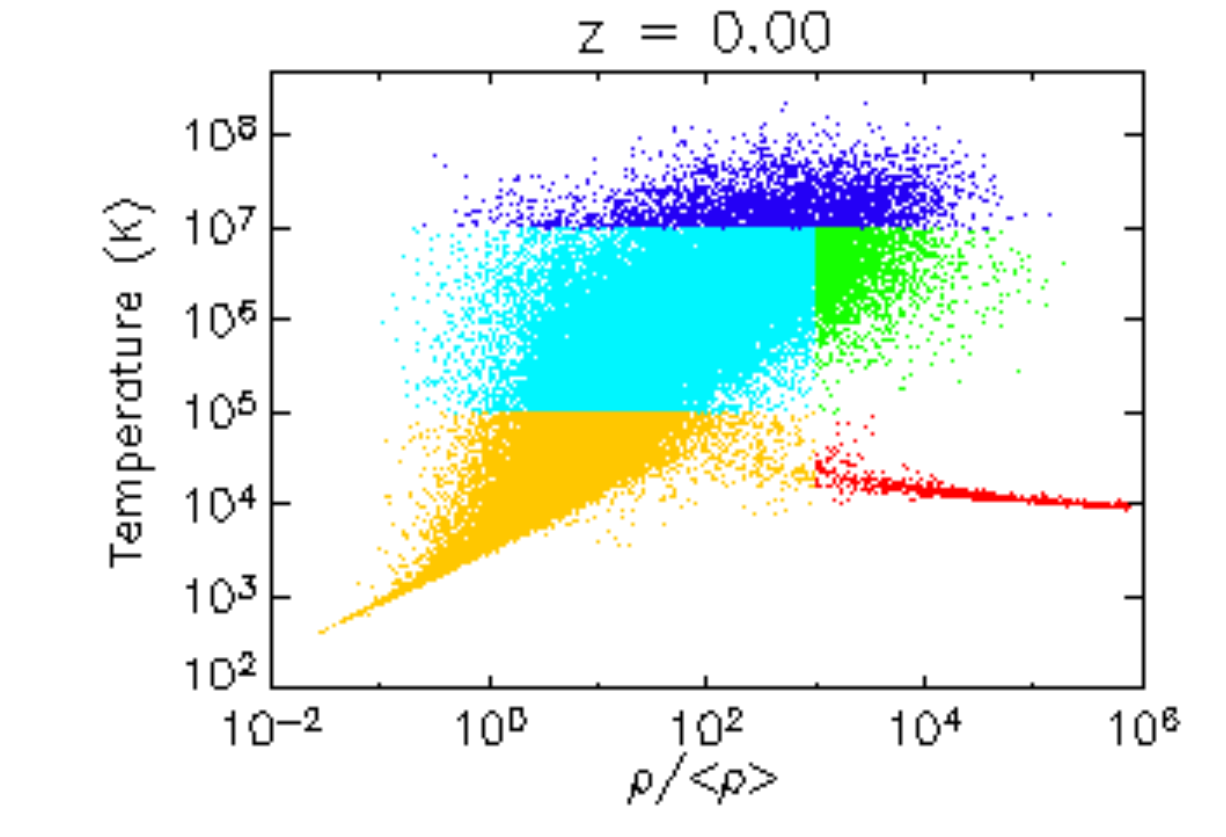}
\caption{Phase diagram of the gas in the temperature-density space. 
Star formation is in \emph{red}, the cold diffuse gas is in \emph{yellow}, 
the WHIM is in \emph{light blue}, groups of galaxies are in \emph{green}, 
and galaxy clusters are in \emph{blue}.}
\end{center}
\label{temp-vs-dens-102-dif}
\end{figure}

From the analysis of the baryonic matter we decided to group the gas 
particles in five phases, depending on density and temperature. At 
temperatures below $10^5$~K we defined two phases, a dense (overdensity 
$\rho  / \langle\rho\rangle > 1000$) cold phase associated to matter 
undergoing star formation, and a cold low density 
($\rho  / \langle\rho\rangle < 1000$) gas that corresponds to the 
diffuse gas that resides in the voids of the web structure of the 
Universe, and that has been identified with the Ly~$\alpha$ 
absorbers. For intermediate 
temperatures ($10^5 < \textrm{T} < 10^7$~K) we have the warm hot gas. 
Also in this case we defined two phases, the low density one 
($\rho  / \langle\rho\rangle < 1000$) that roughly corresponds to 
the filamentary structure of the WHIM, and the high density 
($\rho  / \langle\rho\rangle > 1000$) gas, associated with groups of 
galaxies. Finally, at $\textrm{T} > 10^7$~K, we identify clusters of 
galaxies and the hot intracluster medium (ICM). 
Fig.~\ref{temp-vs-dens-102-dif} shows the phase diagram in the 
temperature-density space.
The choice of the threshold between diffuse and dense WHIM 
is somewhat arbitrary and different authors have used different
thresholds in the range $100<\rho  / \langle\rho\rangle < 1000$.
For consistency with a significant fraction of the literature,
including our previous work (Ursino \& Galeazzi 2006), we adopted 
the definition of Cen \& Ostriker (1999b).

\section{The Simulation}
\label{Simulation}

\begin{figure}
\begin{center}
\includegraphics[width=16cm]{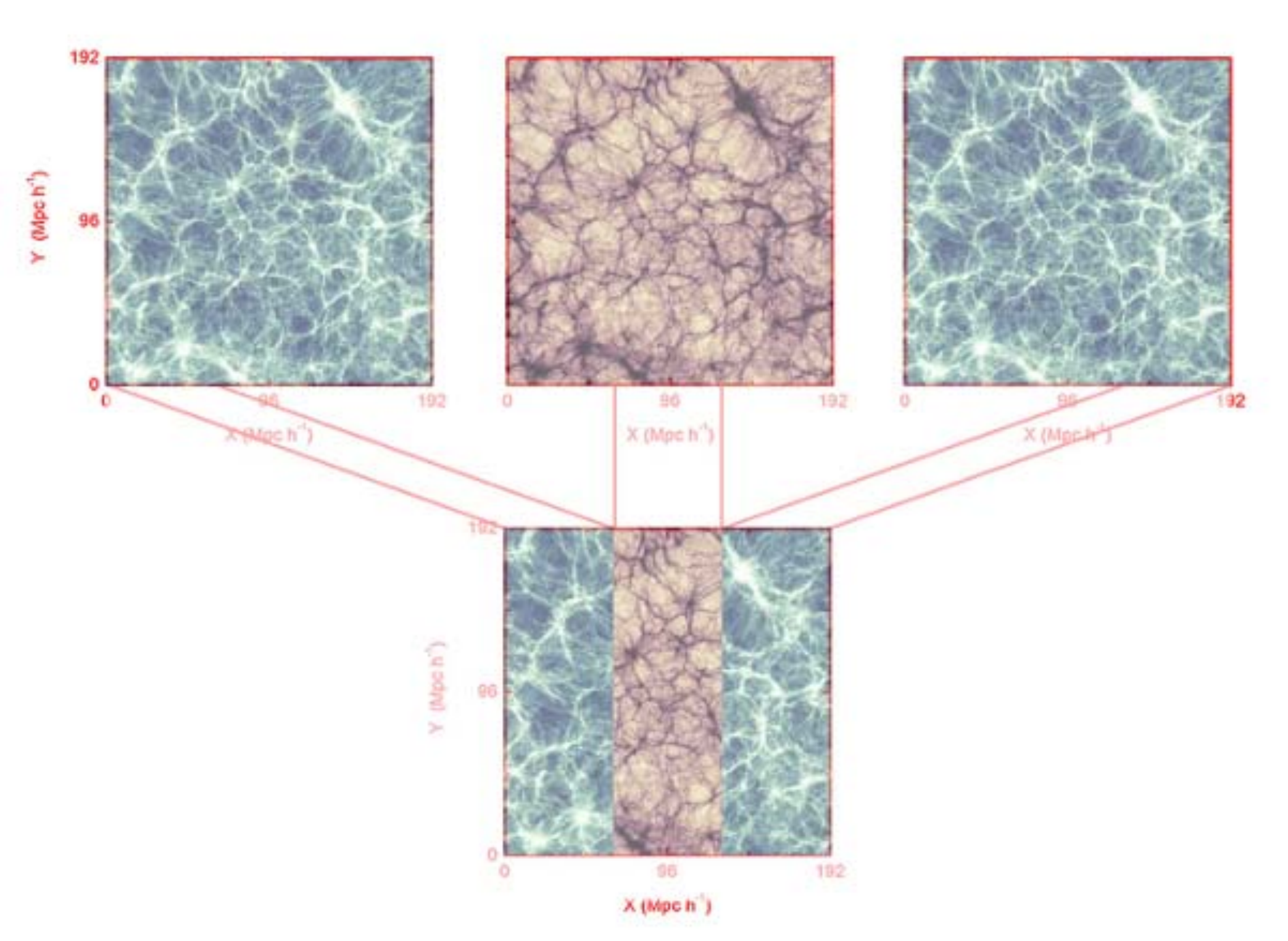}
\caption{The program has redshift intervals too small to contain a 
full snapshot. Three slices of consecutive snapshots are added to form 
a cube for the simulation. The numbers are the redshifts of the 
snapshots used. The figure is elaborated from the map of gas density 
of Fig.~1 in Borgani et al. (2004). The color scale is in arbitrary 
units as the figure is meant only for describing the procedure.}
\end{center}
\label{silicone-slice}
\end{figure}

To create a simulated light cone up to redshift 2, we adopted a procedure 
similar to the one described in Roncarelli et al. (2006), where several 
boxes from the output snapshots of the Borgani simulation were piled 
one after the other. 
Since the boxes have side of 192~h$^{-1}$~Mpc, the redshift interval 
covered by each box is $\Delta z_s\gtrsim0.06$ (increasing at higher redshift). 
The time-step between two consecutive snapshots of the 
Borgani simulation corresponds to $\Delta z_t\gtrsim0.02$ at small 
redshift, increasing at higher redshift, therefore $\Delta z_s$ corresponds to 
the time-step between three consecutive snapshots. For our simulation we 
created one box taking slices of 64~h$^{-1}$~Mpc from three consecutive 
snapshots as shown in Fig.~\ref{silicone-slice}. This allows to 
better follow the evolution of the gas. Since consecutive boxes represent 
the same structure evolved in time and they do not change much for 
small redshift intervals, particular attention was given to avoiding 
encountering the same structures at different redshifts. We therefore used 
random rotations around the line of sight axis, random permutations of 
the coordinate axes of each composite box, and random shifts along the 
three axes. These randomizations give a very large number of degrees 
of freedom to avoid periodicity. The downside of this procedure is 
that it introduces discontinuities between two adjacent snapshots. 
Although these discontinuities are unphysical, their effect on our 
simulation is minor due to the very small probability of finding a
significant WHIM structure right at the edge of the SPH box. 
 
\begin{figure}
\begin{center}
\includegraphics[width=16cm]{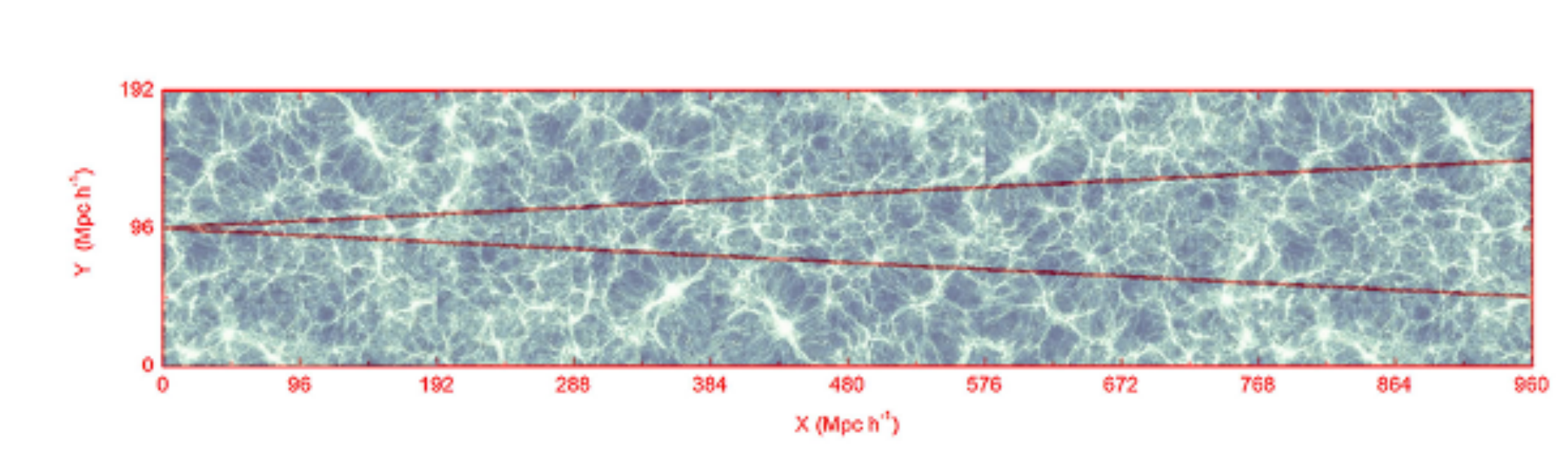}
\caption{Piling up the boxes from Borgani model, the program creates a simulated 
light cone. Particles inside the field of view (defined by the red lines) 
are selected to simulate the X-ray emission. The figure includes the 
first five boxes (discuntinuity at the boundaries are visible), spanning
the redshift interval $0 - \sim0.3$. The color scale is in arbitrary 
units as the figure is meant only for describing the procedure.}
\end{center}
\label{silicone-fov}
\end{figure}

The last step in the simulation is piling up the modified cubes to form a simulated 
light cone. Fig.~\ref{silicone-fov} shows a representation of a simulated 
light cone, with superposed a selected field of view. The code itself 
is capable of spanning any distance as long as the angle of the field 
of view is smaller than the angular size of the last box, but we chose 
a maximum corresponding to redshift $z = 2$, equivalent to 19 cubes. 
Emission from greater distances is expected to be a small fraction of 
the total emission, while at the same time the calculation  drastically 
increases the computing time. 
The angular size of the furthest box, which determines to largest field 
of view that can be contructed in a sigle simulation, 
is $\sim2.1^\circ\times2.1^\circ$. The typical field of view used is 
$1^\circ\times1^\circ$.

The field of view starts from the center at the bottom of the first 
box, and only particles that are inside such field of view (see 
Fig.~\ref{silicone-fov}) are selected. Besides this geometrical 
filter, we adopted two more filters. Since baryons at $T<10^5$~K 
do not emit any significant amount of X-rays, we filtered out all 
the particles with temperature lower than this limit to speed up the 
computational process. At $z=0$ the diffuse warm gas and the star 
forming gas cover $\sim 50$\% of the gas mass (roughly $\sim 50$\% of 
the particles) and that this fraction increases to $\sim 75$\% at $z=2$.
Our filter therefore reduces the computational time by more than a factor of 2. 
One more final filter discriminates between particles of the three 
remaining warm-hot phases and labels them accordingly.

For temperatures in the range $10^5<T<10^7$~K the two main sources 
of radiation between 0.1 and 1~keV are thermal bremsstrahlung and 
emission lines from highly ionized elements (i.e. C\elem{V}, C\elem{VI}, 
O\elem{VII}, O\elem{VIII}, Ne\elem{IX}, Mg\elem{XI}, and Fe\elem{XVII}), 
although our investigation is focused on O\elem{VII} and O\elem{VIII}. 
In order to properly evaluate emission from the lines, it is necessary 
to set metal abundances. 

As we stated previously, we designed the program with the capability 
of working with different metallicity models, in addition from the original 
metallicity  predicted by the hydrodynamic code. This choice is aimed at 
overcoming the lack of precise information about metal formation and 
metal diffusion in the intergalactic medium, and to compare 
the x-ray emission expected from different metal distributions.

The first metallicity model (from now on defined as ``Borgani model'') 
we adopted is the original model coming from the Borgani (2004) 
simulation. This model underestimates the diffusion of metals in most 
of the WHIM, assigning metals to only a fraction of the gas particles. 
The reason of the uneven distribution of metals in the WHIM  
depends on the treatment of metal diffusion in SPH simulations, as 
it is well explained in Fig.~4 of Wiersma et al. (2009b). An SPH 
star-forming particle enriches the neighboring particles and they 
are driven away by the energy released by stars in the star particle. 
The metal enriched particles mix with metal-free gas particles, but 
do not enrich them since there is no diffusion from gas particles. 
Therefore the metal distribution is not evenly smoothed, but 
concentrated in the few gas particles that have been directly enriched 
by star-forming particles. As a result most of the WHIM is poor in 
metals and its emission is underestimated. This can be avoided if a model 
of diffusion of metals from one particle to the others is included 
in the simulation, which is not the case of the simulation we used. 

\begin{figure}
\begin{center}
\includegraphics{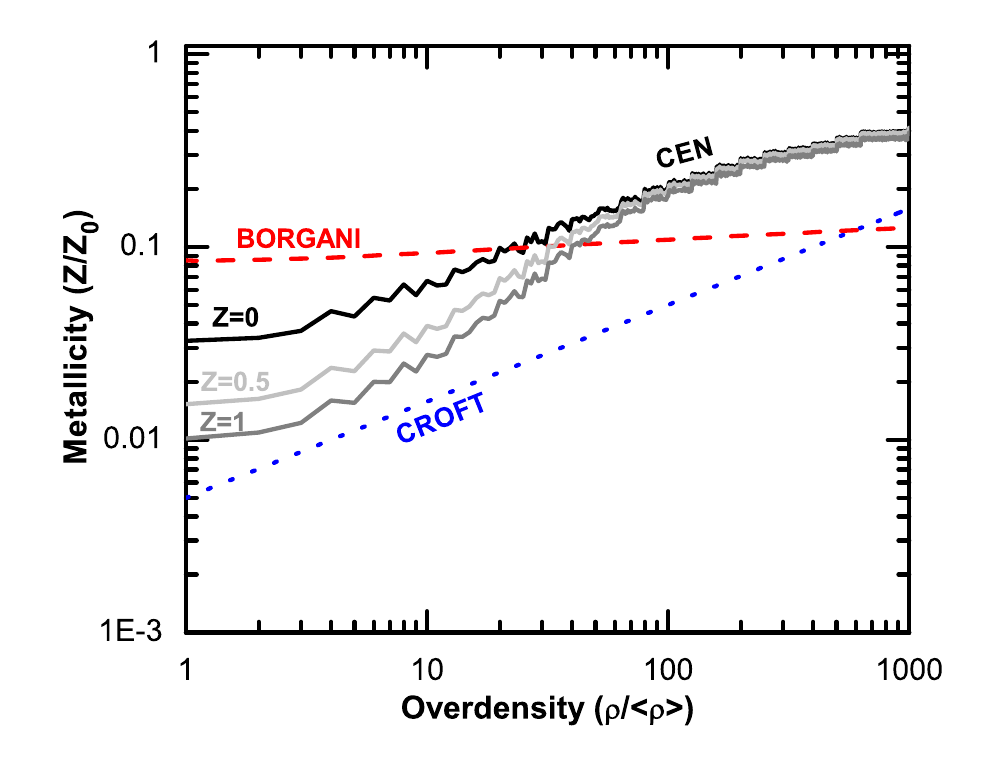}
\caption{Average metallicity as a function of density for the four different 
models used in this paper. 
The Borgani model (\emph{red dashed}) comes directlky from the hydrodynamic simulation 
used in this paper \cite{Borgani04}. The Croft 
model (\emph{blue dotted}) is derived from Croft et al. (2001). The Scatter and Cen models 
(\emph{black solid}) come from a random selection based on the distribution function from 
Cen cubes \cite{CenOst99}. The two models are identical at z=0, but the 
Cen model includes a dependence on redshift following Cen \& Ostriker (1999b); 
the average metallicities for the Cen model at z=0.5, and z=1 is also shown 
(\emph{gray solid}).}
\end{center}
\label{metallicity-vs-density}
\end{figure}

\label{metal-models}
In addition to the metallicity extracted from the Borgani model, we 
used three analytical models based on relations between metallicity 
and density (see Fig. \ref{metallicity-vs-density}). 

The first analytical model (defined as ``Croft model'') links the 
metallicity directly to the gas density, using the relation 
$Z\propto(\rho/\overline{\rho})^{1/2}$. 
Metallicity is normalized to $Z=0.005$~$Z\odot$ at $\rho=\overline\rho$, 
so that it matches the measured metallicity of the Ly-$\alpha$ forest, 
while an upper limit of $Z=0.3$~$Z_\odot$ fits well with data on clusters 
\cite{Fang05}. This model is in agreement with the metallicity predicted 
by Cen \& Ostriker (1999a) at $z=3$ and is about a factor 5 lower than 
the IGM metallicity at more recent times. Nevertheless we adopted this 
model as comparison since it has also been used by other authors 
\cite{Croft01, Fang05}. 

The  second analytical metallicity model (which we called the 
``Scatter model'') is based on the distribution function of metallicity 
from Cen and Ostriker (1999b) at redshift $z=0$. Part of the output 
of that simulation consists of three boxes of $512^3$ cells with 
values of temperature, density, and metallicity. Using the three 
boxes we generated the probability distribution function of 
metallicity as a function of density, where metallicity is divided 
in 110 intervals from $10^{-7}$ to $10^4$~$Z_\odot$ and density is 
divided in 70 intervals from $10^{-3}$ to $10^4$~$\rho_b$. When this 
model is selected, the code reads the density of each particle and assigns 
a random metallicity based on the probability distribution function 
at the corresponding density. The average distribution of metallicity 
for this model is represented by the black curve in 
Fig.~\ref{metallicity-vs-density}. This model has the highest 
metallicity among the four models (a factor $2-3$ for overdensities 
between 10 and 1000) and, since at first order the intensity of the 
lines depends linearly on metallicity, we use the emission with the 
scatter model as an upper limit for our set of simulations. 

The third analytical metallicity model, labeled ``Cen model'', 
is an improved version of the scatter model, where we also include 
redshift dependence evaluated from Fig.~2 of Cen \& Ostriker (1999b). 
A ``random'' metallicity is initially evaluated following the same
procedure used for the scatter model, then it is modified according 
to redshift of the particle. We note that the redshift dependence of
the metallicity is optimized for WHIM particles and would overestimate 
the metallicity of clusters and groups.
By definition the average values of the Cen model at z=0 are identical
to those of the scatter model. To show the redshift variation, in 
Fig.~\ref{metallicity-vs-density} we plot the average values of the 
Cen model at $z=0.5$, and $z=1$ in tones of gray.
We introduced this model for two main reasons: to evaluate the 
influence of time evolution on metallicity and to compare the 
predictions of this model with the results of our previous work 
\cite{Ursino06}, where we used the same fitting function to estimate 
time evolution of metallicity.

We stress the fact that in the original Borgani simulation the cooling 
function does not depend on metals but only on primordial abundances 
and that the models of metallicity that we adopt, are introduced 
\emph{a posteriori} and do not affect the baryon history. It has been 
shown that including a self consistent metal dependence in a 
simulation increases the cooling rates, resulting in a lower emission 
(above all for strongly emitting gas) than what is predicted using 
only primordial abundances, even by an order of magnitude 
\cite{Bertone09}.

The following step is to calculate the spectrum of emitted photons, 
as a function of electron 
number density $n_e$, temperature $T$ and metallicity $Z$ for every 
SPH particle. We did not consider peculiar velocities. Velocities of 
a few hundred km~s$^{-1}$ correspond to $\Delta z\sim0.001$, more 
than a factor 30 less compared to the redshift interval between two 
boxes. These velocities correspond to displacements of the emitters 
of less than 10~Mpc, if we attempt to estimate the distance from the 
redshifted spectrum of an absorber. We used the XSPEC version of the 
APEC model\footnote{http://heasarc.gsfc.nasa.gov/docs/xanadu/xspec/} 
\cite{Smith01} using solar relative abundances to produce the spectra, 
in the assumption of an optically thin gas in CIE and no 
photo-ionization from a background radiation. We assumed CIE since we 
expect to probe the WHIM mainly in regions at higher temperature and 
density, where the ionization balance is dominated by collisions. 
Models show that the WHIM could depart from ionization equilibrium  
\cite{Yoshida05, CenFan06} and the abundances of ions in the case of 
non-equilibrium are generally higher \cite{Gnat07}, however the 
expected differences between the observables in ionization equilibrium 
and non-equilibrium are small \cite{Yoshikawa06}. In details, our 
procedure goes as follows. We created a grid of spectra as a function 
of temperature and 
metallicity, equally spaced on a logarithmic scale both in temperature 
between $10^5$~K and $10^8$~K, and metallicity Z between $5\times10^{-4}$~Z$_\odot$
and $5$~Z$_\odot$ respectively. We set the energy resolution at 1~eV 
between 0.05 and 3~keV, 
and at 50~eV between 3 and 50~keV. We interpolated from the grid, using 
the temperature and metallicity of a given SPH particle to compute a 
reference spectrum $\Lambda_{T,Z}(E_0)$, with $E_0$ the rest-frame 
energy. Then we calculated the emitted spectrum, corresponding to the 
number of emitted photons per unit of time, with the formula
\begin{equation}
s_0(E_0) = x_e n_{\rm H}^2  V \Lambda_{T,Z}(E_0) \ ,
\end{equation}
where $V$ is the physical volume of the particle (defined as mass of 
the particle divided by density), $n_{\rm H}$ is the number density of 
hydrogen nuclei (given by the SPH simulation), and the density of electrons
is given by $n_e = x_e \times n_{\rm H}$, with $x_e=1.225$ 
(assuming near full ionization).

Finally we computed the observer-frame spectrum $s(E)$, applying the energy redshift
$E=E_0(1+z)$ to $s_0(E_0)$, and the corresponding flux in each energy channel
\begin{equation}
\label{eq:sb}
S(E) = \frac{s(E)}{4 \pi d_{\rm c}^2 (1+z)} \ ,
\end{equation}
where $d_{\rm c}$ is the comoving distance of the particle from the observer
and the factor $(1+z)$ accounts for the redshift dimming of the signal.

This quantity is then distributed in the different map pixels by using the
SPH smoothing kernel
\begin{equation}
w(r) \propto \left\{ \begin{array}{ll}
1-6r^2+6r^3, & 0\le r \le 0.5 \nonumber \\
2(1-r)^3, & 0.5\le r \le 1 \\
0, & r \ge 1 \ , \nonumber
\end{array} \right.
\label{eq:kernel}
\end{equation}
where $r\equiv\Delta\theta/\alpha_h$ is the angular
distance from the particle position in the map, in units of the angle $\alpha_h$
subtended by the particle smoothing length provided by the hydrodynamic
code.
In this kind of computation the smoothing procedure results to be the
most time consuming. However the mathematical properties of the smoothing
function can help speeding up the computation. First of all,
since the function of eq.~\ref{eq:kernel} is the approximation of a
2-dimensional Gaussian in the sky plane, the distribution can be separated
into the product of its two components along the map axis,
$w(r) \simeq w(x)\times w(y)$. Then, since the function is integrable, we can
directly define $W(x) \equiv \int_{-1}^x w(\tilde{x}) d\tilde{x}$ and 
use it for the computation. The normalization of $w(x)$ is chosen to 
have $W(1)=1$. Therefore, for a given pixel $(i,j)$ we compute its 
corresponding flux fraction
\begin{equation}
f_{i,j} = [W(x_1)-W(x_0)] \times [W(y_1)-W(y_0)] \ ,
\end{equation}
where $(x_0,y_0,x_1,y_1)$ identify the pixel limits in the two
directions. We highlight also the fact that our definition of $W(x)$ ensures
that $\Sigma_{i,j} f_{i,j} = 1$ without the need to renormalize it.

This procedure is done separately for the three gas phases (WHIM, dense WHIM,
and hot gas) in order to keep the information about their different
contributions. 

The final result of the simulations is a three dimensional array (the two angular 
coordinates and the energy) for each gas phase (Hot, WHIM, 
and Dense WHIM) containing photon counts per second per cm$^2$. 
We also saved one array for each 
box used to build the simulated light cone, so that we have information about 
the spectrum as a function of redshift. 

At the end of the process  
we included the absorption due to the neutral hydrogen in our 
galaxy using the model described by Morrison and McCammon (1983), with a 
typical value at high latitude of nH 
$=1.8\times10^{20}$~cm$^{-2}$ \cite{McCammon02}.

\section{Results}
\label{Results}

The imaging capability of the simulation gives us the opportunity to study the WHIM 
morphology and emission, or focus on the properties of individual objects. 
We can also extract energy spectra for every pixel of the image. 

\begin{figure}
\begin{center}
\includegraphics{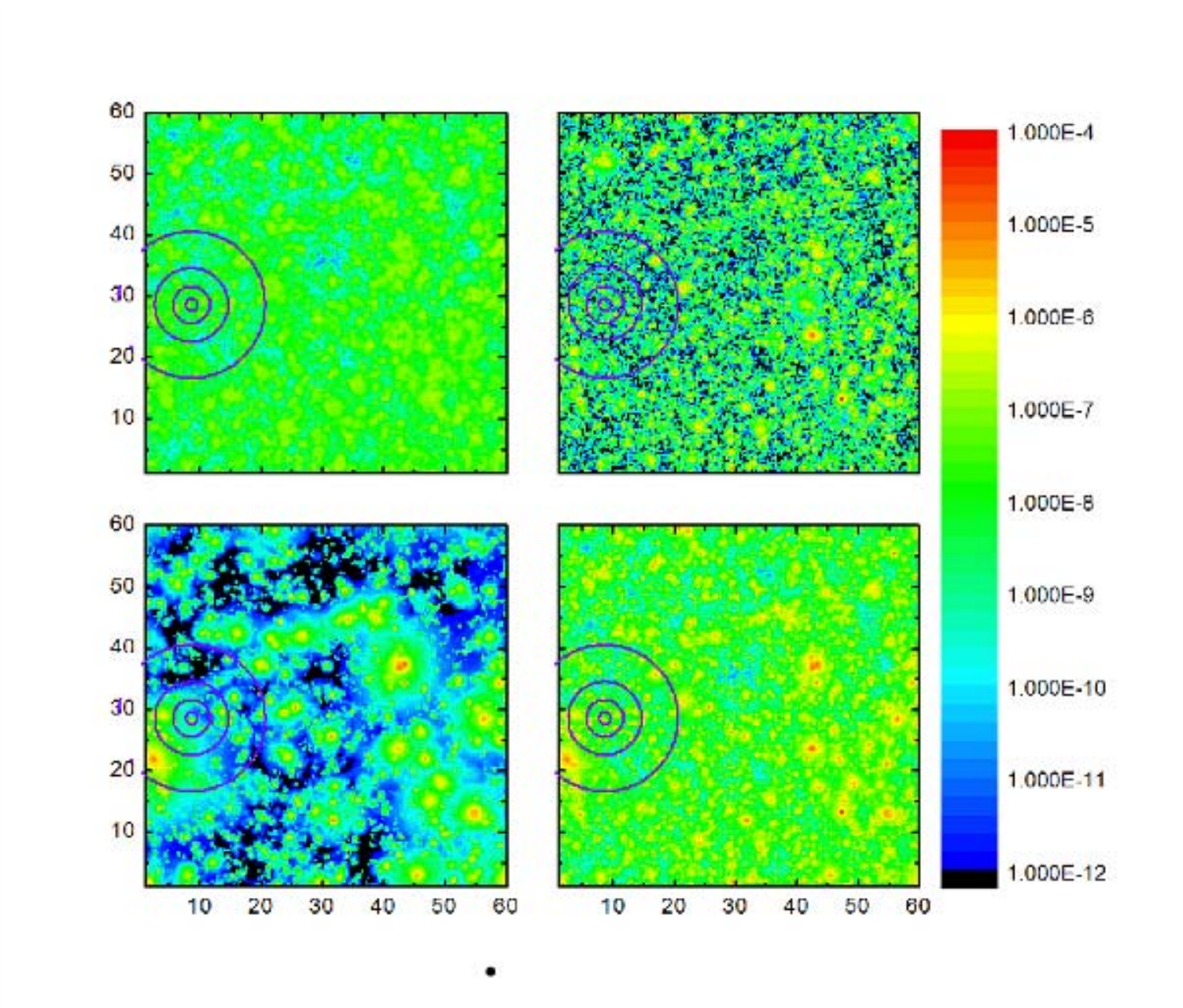}
\caption{Simulations of the emitted X-rays from the WHIM 
(\emph{top left}), the dense WHIM (\emph{top right}), the hot gas 
(\emph{bottom left}), and the total gas (\emph{bottom right}) in the 
full line of sight, up to $z=2$. 
The circles represent different fields of 
view with radii 1', 3', 6', and 12'. The map has a FOV of 
$1^\circ\times1^\circ$ and $256\times256$ pixels (resolution of 14'') 
in the energy range $380 - 950$~eV, and uses the Croft metallicity 
model. Units for the color bar are photons~cm$^{-2}$~s$^{-1}$.}
\end{center}
\label{final_um_060315a_Map_dif_tot}
\end{figure}

\begin{figure}
\begin{center}
\includegraphics{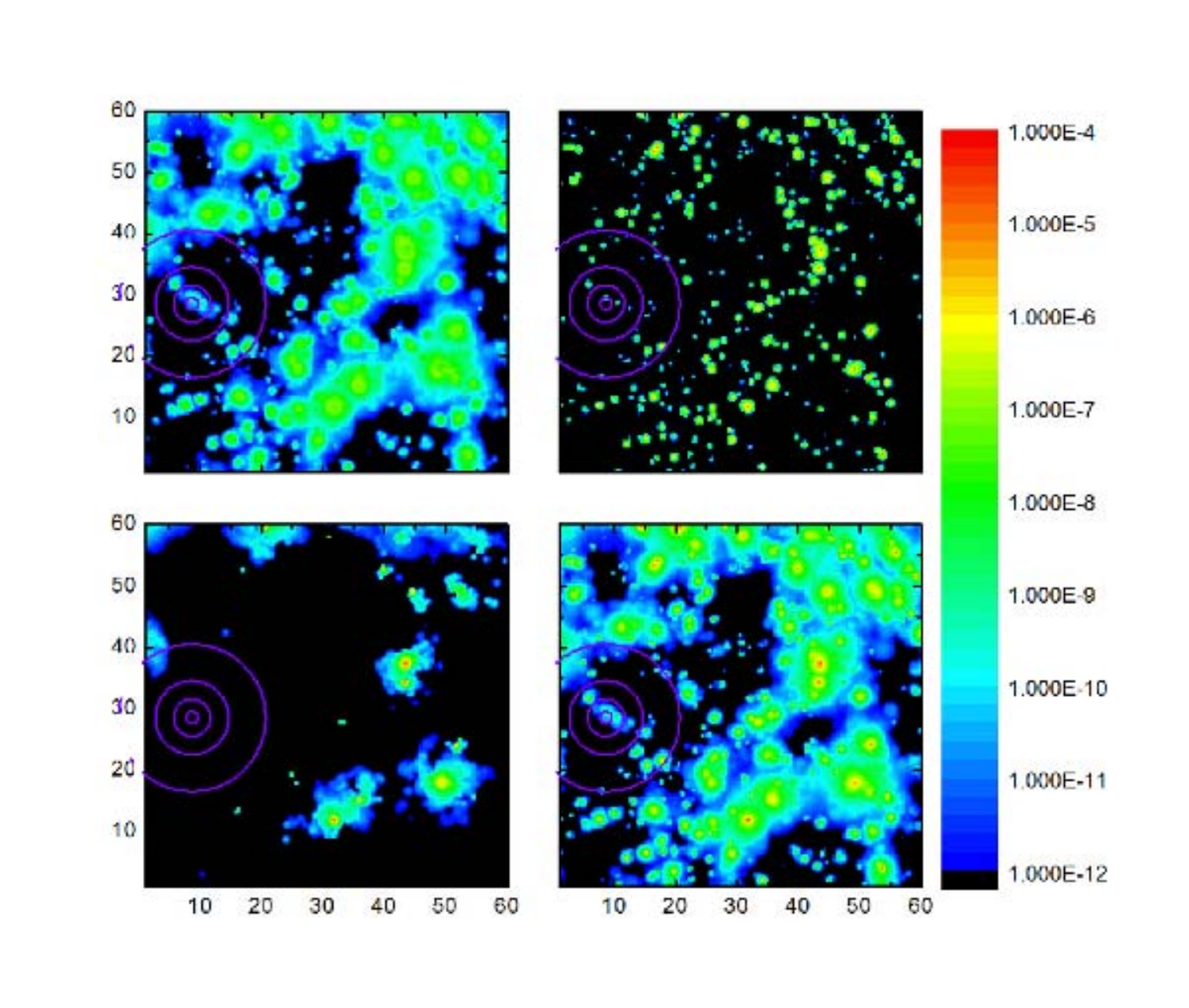}
\caption{Simulations of the emitted X-rays from the WHIM 
(\emph{top left}), the dense WHIM (\emph{top right}), the hot gas 
(\emph{bottom left}), and the total gas (\emph{bottom right}) 
from slice \# 4 ($z\sim0.25$, thickness 192~h$^{-1}$~Mpc) 
of the same simulation of Fig.~\ref{final_um_060315a_Map_dif_tot}. 
The circles represent different fields of 
view with radii 1', 3', 6', and 12'. The map has a FOV of 
$1^\circ\times1^\circ$ and $256\times256$ pixels (resolution of 14'')  
in the energy range $380 - 950$~eV, and uses the Croft metallicity 
model. Units for the color bar are photons~cm$^{-2}$~s$^{-1}$.}
\end{center}
\label{final_um_060315a_Map_dif_04}
\end{figure}

As an example of the capabilities of our simulations, Figs.~\ref{final_um_060315a_Map_dif_tot} 
and \ref{final_um_060315a_Map_dif_04} show images in the energy band 
$380 - 950$~eV, with a field of view of $1^\circ\times1^\circ$ and $256\times256$ 
pixels, with an angular resolution of $14''\times14''$. The simulation runs 
to a distance equivalent to $z=2$.

The energy band $380 - 950$~eV is adopted throughout all the paper, 
unless otherwise stated, and  is the same as the one we used in our 
previous work \cite{Ursino06}. We chose the upper limit to include 
the emission from Ne\elem{IX} at 921~eV and the strong Fe\elem{XVII} 
lines in the energy band $0.725 - 0.827$~keV, while with the lower 
we could exclude the strong C\elem{V} and C\elem{VI} lines at 0.308~keV 
and 0.367~keV respectively. This makes us sensitive to 
O\elem{VII} out to redshifts of z=0.5. It is also advantageous 
to avoid instrumental effects due to the neutral carbon absorption 
edge at 0.284~keV, which is present in most instruments. 

The angular resolution is a trade-off between the requirements of 
accuracy and computing resources. In our previous work we have shown 
that the detected WHIM emission depends on the angular resolution 
\cite{Ursino06}. We have seen that the characteristic angular size of 
filaments is of the order of a few arcminutes and that with a coarse 
resolution the probability of detecting individual WHIM filaments 
becomes negligible.
Similarly, a more detailed analysis on the effect of the angular 
resolution has also been reported in \cite{Bertone09}. 
The chosen angular resolution is well below the typical WHIM scale
as indicated in both papers, and should allow a good, unbiased 
characterization of the angular distribution of the WHIM emission.

\begin{figure}
\begin{center}
\includegraphics{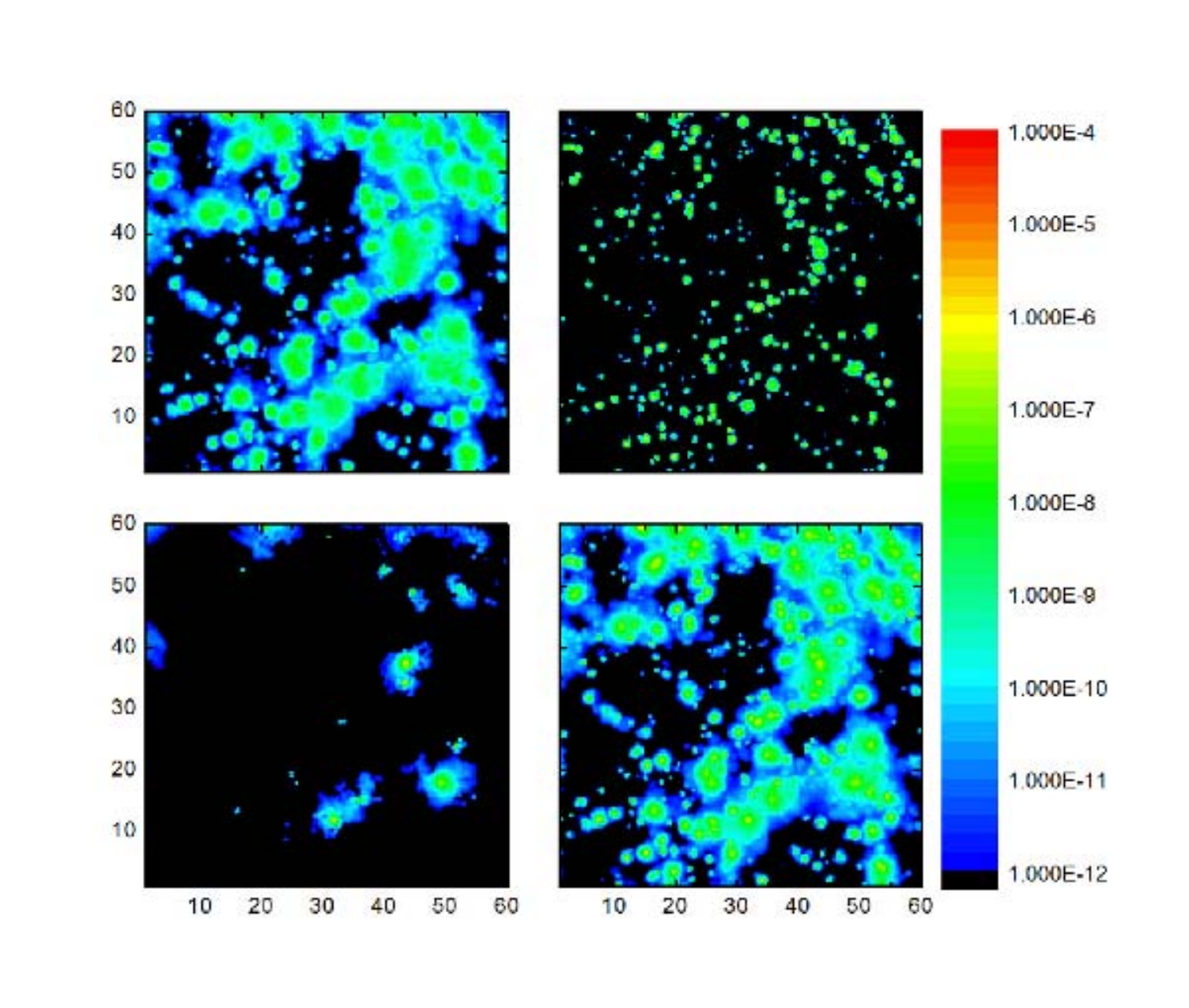}
\caption{Simulations of the emitted X-rays from the WHIM 
(\emph{top}), the dense WHIM (\emph{center}), and the hot gas 
(\emph{bottom})  
in the same field of view and redshift range of the simulation of 
Fig.~\ref{final_um_060315a_Map_dif_04}, the energy range 
corresponds to the O\elem{VII} ($561-574$~eV) band redshifted at z=0.25. 
The map has a FOV of 
$1^\circ\times1^\circ$ and $256\times256$ pixels (resolution of 14'') and 
uses the Croft metallicity model. Units for the color bar are 
photons~cm$^{-2}$~s$^{-1}$.}
\end{center}
\label{final_um_060315a_Map_dif_04_OVII}
\end{figure}

For this exercise we generated images for each of the three gas phases, 
for each redshift slice corresponding to a box of the hydrodynamic 
simulation,and for the full line of sight, and we selected a region 
in the image to extract the relative spectra for different circular 
FoVs. Looking at the images of the slice in the redshift interval 
$0.201-0.273$ (Fig.~\ref{final_um_060315a_Map_dif_04}), the different 
nature of the three phases is clear. The WHIM is collected in large, 
diffuse regions and is sparse throughout the whole image in 
filaments. The dense WHIM 
is grouped in much more compact objects, and is found in the inner 
part of the WHIM regions, where it traces the WHIM filaments. 
The hot gas is found in few large objects, close to the regions where 
the WHIM emission is stronger. The general picture confirms the 
assumption that was made when we defined the three phases. The hot 
gas is associated to the big clusters of galaxies while the dense 
WHIM corresponds to the groups of galaxies. The diffuse WHIM, on the 
other side, appears mostly as haloes that envelope the groups. 
As it should be expected, when the contribution at 
all redshifts is included, the filamentary structure of the WHIM 
is partially hidden in these images 
(Fig.~\ref{final_um_060315a_Map_dif_tot}). This is due to the large 
energy interval used in the generation of these images. However, 
when the energy information is taken into account by focusing for 
example on a narrow energy interval containing the redshifted 
O\elem{VII} line, the spatial information about the three phases can 
still be extracted. Figure \ref{final_um_060315a_Map_dif_04_OVII} 
shows the image obtained by generating an image of the full line of 
sight for thenarrow energy band corresponding to the energy of 
O\elem{VII} lines with redshift between $0.201$ and $0.273$. Notice 
that we obtain the same spatial information of 
Fig.~\ref{final_um_060315a_Map_dif_04} for the WHIM, while the hot 
and dense sources are more compact than for the full energy band. In 
the case of dense sources this indicates that we see O\elem{VII} 
emission mostly from their core. With the hot sources, on the other 
side, although we are indeed probing the O\elem{VII} band, we 
actually see photons emitted via bremsstrahlung, which dominates (in 
this band) at high temperatures. 

\begin{figure}
\begin{center}
\includegraphics[width=8cm]{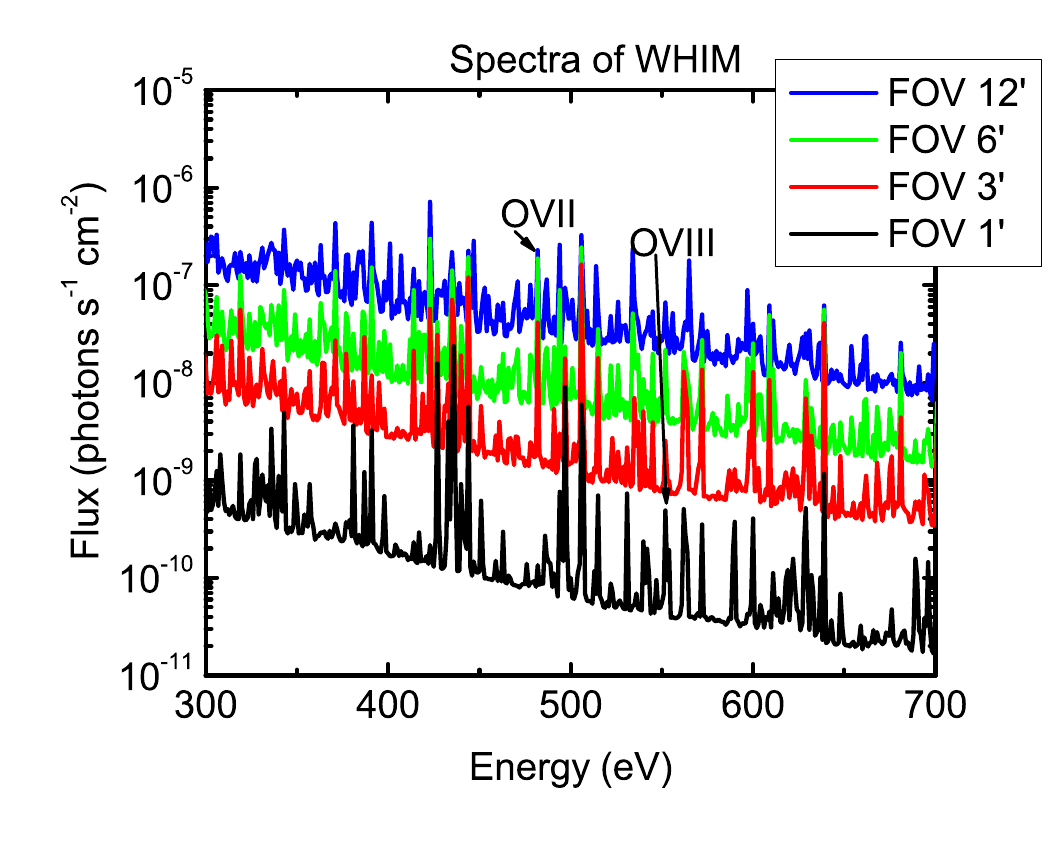}
\includegraphics[width=8cm]{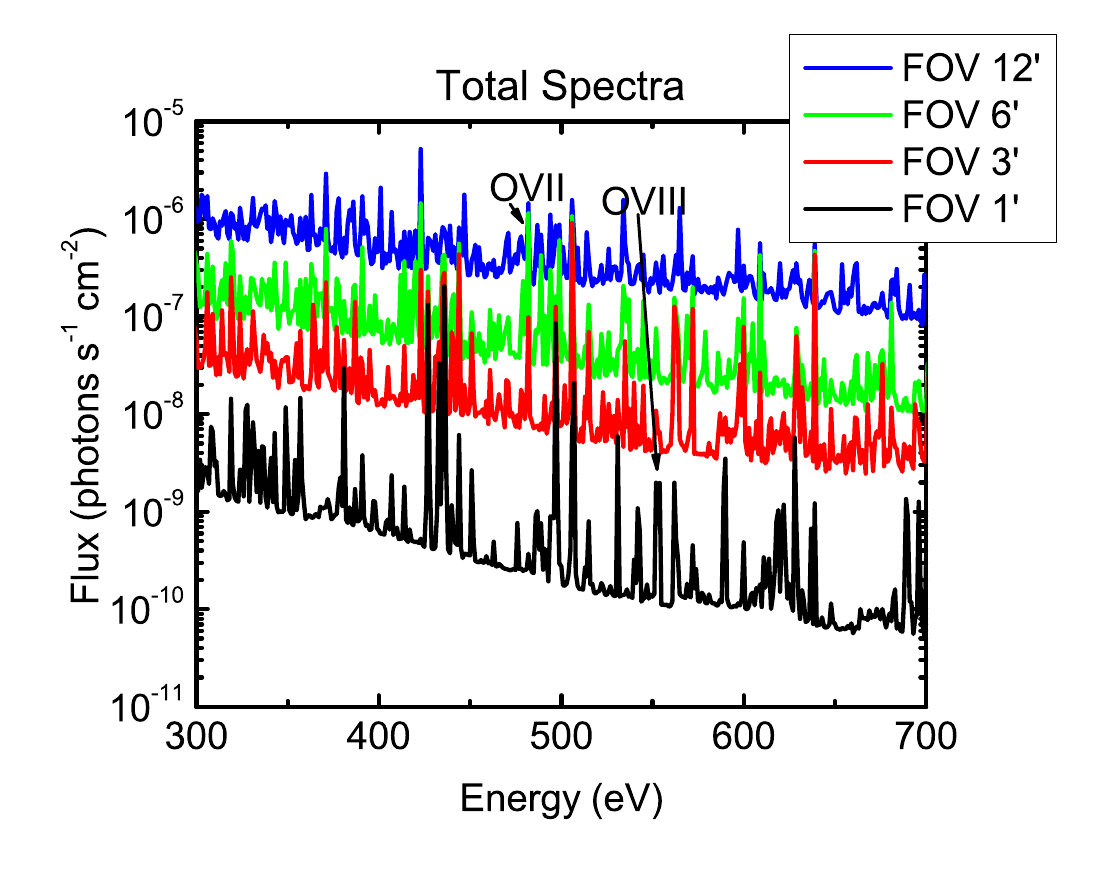}
\caption{Spectra extracted from the pointing shown in Fig. 
\ref{final_um_060315a_Map_dif_04} 
with radii 1', 3', 6', and 12', using data for the WHIM only 
(\emph{left}) and for the sum of WHIM, dense WHIM, and ICM 
(\emph{right}). }
\end{center}
\label{silicone_spectra_whim}
\end{figure}

Figure \ref{silicone_spectra_whim} shows the spectra extracted from 
the selected fields of view in figures 
\ref{final_um_060315a_Map_dif_tot} and 
\ref{final_um_060315a_Map_dif_04}, along the full line of sight. We 
can use it to set a limit on the angular resolution required to study 
the WHIM. The 1' field of view is centered on a WHIM filament. The 
spectrum shows O\elem{VIII} emission lines from the selected filament, 
together with lines emitted by baryons at different redshifts. When 
we increase the field of view to 3', we also find the O\elem{VII} 
line from a different region of the filament. At the same time the 
overall flux from the WHIM at all redshifts increases due to the 
contribution of other regions, making it more difficult to identify 
the signal of the filament being studied. At even larger fields of 
view, the lines from the filament are overshadowed by the total WHIM 
flux. This indicates that, at redshift $z\sim0.25$, an angular 
resolution of a few arcminutes is necessary in order to detect a 
filament. A more systematic investigation of the angular distribution 
of the WHIM emission is discussed in the next section.

The spectrum on the right side of Figure~\ref{silicone_spectra_whim} 
shows that the emission of the dense WHIM and the hot gas is stronger 
than the WHIM, and the flux is between 5 and 10 times higher. It is 
important to note that the cluster contribution to the spectrum in 
the soft X-rays is mostly due to bremsstrahlung.To separate the WHIM 
from these two phases we need to identify the emission lines, and this 
sets the requirement on the energy resolution of any instrument 
designed to study the WHIM. An energy resolution of a few eVs is 
necessary to resolve single lines, in particular if we want to 
identify the O\elem{VII} triplet at $\sim 570$~eV, the main tracer 
of the WHIM. 

\section{X-ray Emission and Metallicity}
\label{Metallicity-emission}
 
We generated four sets of identical simulations, with the metallicity 
as the only parameter that was changed, using the models described in 
section \ref{metal-models}. For simplicity we refer to those models as 
``Borgani'', ``Croft'', ``Scatter'', and ``Cen''.

To evaluate the contribution of the WHIM to the total diffuse X-ray 
emission we compared our simulations with results from the 
X-ray Quantum Calorimeter (XQC) sounding rocket program (McCammon 
et al. 2002) and the $ROSAT$ All Sky Survey (RASS - Snowdnen et al. 1994).
XQC is the only current mission using high resolution 
microcalorimeters for the study of the DXB in the energy range 
50-2000 eV, while RASS is the most accurate X-ray survey below 
1~keV and are considered the benchmarks for any soft DXB studies. 
The most recent XQC published results (McCammon et al. 2002) predict 
a surface brightness of 29.3\phot~ in the 0.380-0.950~keV energy 
band. In table \ref{surf-bright} we report the surface brightness 
predicted by the four simulations in the same energy band and the 
predicted flux in the RASS $R4$ and $R5$ bands (Snowdnen et al. 1994) in units of 
10$^{-6}$~photons~s$^{-1}$~arcmin$^{-2}$ (the default RASS units). 
The average RASS flux in the R4+R5 bands is 
$\sim$~139 10$^{-6}$~photons~s$^{-1}$~arcmin$^{-2}$.
The Borgani model has the lowest brightness, $0.77\pm0.04$\phot in the
 0.380-0.950~keV energy range, and can be used to put a lower limit to 
the WHIM emission. We remind that 
even if this model has a high average metallicity at low densities 
(as shown in Fig.~\ref{metallicity-vs-density} and table 
\ref{surf-bright}), most of the particles actually have no metals due 
to the poor diffusion of metals to low density regions in the 
hydrodynamic code. Moreover, the few particles with metals have low 
density and, since emission goes as density squared, they give little 
contribution as well.

The Croft model predicts a relatively low emission as well, $1.84\pm0.08$\phot. 
This model was designed to work at high redshift ($z\sim3$), where 
the gas is still rather poor in metals. Although being somewhat 
unrealistic, this model is a test of intermediate metallicity. 

The Scatter model, on the other hand, was created for gas at $z=0$, 
and slightly overestimates the observed metallicites at $0<z<0.5$ 
($\textrm{Z}\sim 0.3 \textrm{Z}_\odot$ for groups and clusters 
and $\textrm{Z}\sim0.1\textrm{Z}_\odot$ for diffuse gas, as seen in 
\S~\ref{introduction}), but it gives much higher metallicities (possibly 
up to a factor 10) than what is measured at higher redshift ($0.5<z<2$).  
The average surface brightness is $4.3\pm0.2$\phot. This value is the 
highest predicted by the set of simulations and we use this as an upper 
limit to the flux. 

Using the Cen metallicity model, the simulation predicts a surface 
brightness of $4.2\pm0.2$\phot. All the values are smaller than what 
we obtained in our previous work \cite{Ursino06} where, using the 
hydrodynamic model by Cen and Ostriker, (1999a) we obtained a predicted 
brightness of $6.9\pm0.9$\phot. It must be noted that in the previous
work we analytically extrapolated the values of temperature, 
density, and metallicity from the dataset at $z=0$ as it was the only 
dataset available.  It is therefore possible that the evolution of 
the gas is rather different than what is used in the current simulations. 

Focusing on the O\elem{VIII} line at 650~eV in Fig.~15 of McCammon 
et al. (2002) we see that, if we consider the local foreground and 
the AGNs component, there is still room for an extragalactic 
contribution of $\sim9$\phot~kev$^{-1}$ ($\sim0.09$\phot ~if we assume 
an energy interval $\Delta$E$=10$~eV). In the same band, our four 
metallicity models predict a surface brightness of $9.9E-3\pm0.6E-3$\phot, 
$0.025\pm0.002$\phot, $0.059\pm0.004$\phot, and $0.058\pm0.004$\phot 
~respectively, all  within the limits of the X-ray Quantum Calorimeter 
(XQC) sounding rocket data \cite{McCammon02}.

The two higher metallicity models predict values that are half of  
the ROSAT data in Fig.~1 of Kuntz et al. (2001). The absorbed 
extragalactic component in the $380-950$~eV, in fact, has an almost constant 
value of $\sim10$~keV~cm$^{-2}$~s$^{-1}$~sr$^{-1}$~keV$^{-1}$, 
corresponding to $\sim8.8$\phot.

Recently Hickox and Markevitch (2007) quantified the unresolved XRB 
in the Chandra deep fields as $(1.0\pm0.2)
\times10^{-12}$~ergs~cm$^{-2}$~s$^{-1}$~deg$^{-2}$ in the $0.65-1$~keV band. 
Although these measurement were performed in a very limited region of 
the sky (a circle of radius 3.2'), this value constitutes an upper limit to 
the WHIM emission. 
All of our models fall below this limit, predicting, in the same band 
$(6.5\pm0.4)\times10^{-14}$~ergs~cm$^{-2}$~s$^{-1}$~deg$^{-2}$, 
$(1.5\pm0.1)\times10^{-13}$~ergs~cm$^{-2}$~s$^{-1}$~deg$^{-2}$, 
$(3.3\pm0.3)\times10^{-13}$~ergs~cm$^{-2}$~s$^{-1}$~deg$^{-2}$, and 
$(3.2\pm0.3)\times10^{-13}$~ergs~cm$^{-2}$~s$^{-1}$~deg$^{-2}$ 
respectively.

The results obtained with the Borgani model can also be compared with 
Roncarelli et al. (2006), who obtained an estimates on the DXB starting 
from the same cosmological simulation and with a similar method. They 
obtain $5 \times 10^{-13}$~ergs~cm$^{-2}$~s$^{-1}$~deg$^{-2}$ in the 
same band (as extrapolated by Hickox \& Markevitch 2007), thus almost 
an order of magnitude higher than the 
$(6.5\pm0.4)\times10^{-14}$~ergs~cm$^{-2}$~s$^{-1}$~deg$^{-2}$ 
obtained with the Borgani model, 
although within the observational upper limit. This difference is due to 
the fact that Roncarelli et al. (2006) adopted an observationally 
oriented approach by excluding from their maps the extended sources 
detected by Chandra observations. This means that their estimate includes 
also the emission from unresolved clusters and groups (corresponding to 
our hot and dense WHIM phases, respectively) that are instead excluded 
in our estimate. 

Measurements of the autocorrelation function for a set of XMM-Newton 
observations \cite{Galeazzi09} point to the fact that the X-Ray 
emission in the $0.4-0.6$~keV band from the WHIM is $12\pm5\%$ of the 
total extragalactic diffuse emission. Our models predict $10\pm1\%$, 
$12\pm1\%$, $17\pm1\%$, and $13\pm1\%$, in agreement with 
the observational data.

\begin{figure}
\begin{center}
\includegraphics{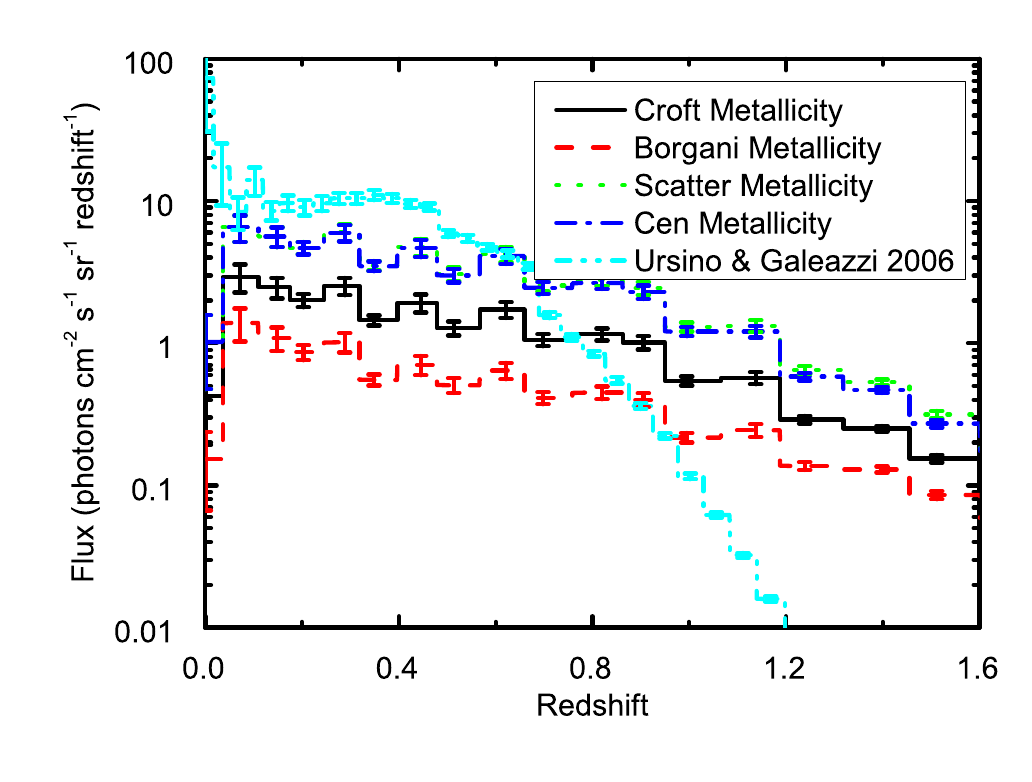}
\caption{Expected surface brightness in the energy range 0.380-0.950 keV due to 
the WHIM as a function of redshift depending on the metal model. Borgani metallicity 
is \emph{dashed red}, Croft metallicity is \emph{solid black}, Scatter metallicity 
is \emph{dotted green}, and Cen metallicity is \emph{dot-dashed blue}. For 
comparison, the results of our previous work \cite{Ursino06} are in 
\emph{double dot-dashed light blue}.}
\end{center}
\label{silicone-fluxredshift-whim}
\end{figure}

In order to understand the dependence of flux from metallicity and 
redshift of the sources, we calculated the average flux of each 
slice of the simulations and compared with the results of our previous 
work \cite{Ursino06}. Figure \ref{silicone-fluxredshift-whim} clearly shows 
that, while the overall photon budget is comparable between the old 
and new analysis, the dependence on redshift behaves in very different 
ways, being much steeper in the case based on Cen \& Ostriker (1999a) 
simulations. Focusing on the data from the current simulation, we see 
that the dependence from redshift changes with metallicity model. 
The four models show the same trend, with a rather slow decrease of 
photon flux at increasing redshift (compared to the older simulations), 
but the relative steepness is different. For the Borgani metallicity, 
photons coming from low redshift are ten times more than those coming 
from very high redshift, in the case of Croft metallicity the ratio of 
photons from near sources over photons from distant sources is around 
20, and for the highest metallicity models this ratio is of the order 
of $\sim30$. Since in our models metallicity depends directly on density, 
the metal abundance is higher in high density regions, where there is 
more star formation. At low redshift, when the Universe is older, there 
are more high density regions compared to the young Universe, and 
therefore there are more high metallicity regions. This makes the 
difference between the abundance of metals at low redshift and at high 
redshift bigger for those models where metallicity is stronger. 
This effect is due only to the relation between density and metallicity, 
and does not depend on any assumption on star formation history. 

\begin{figure}
\begin{center}
\includegraphics{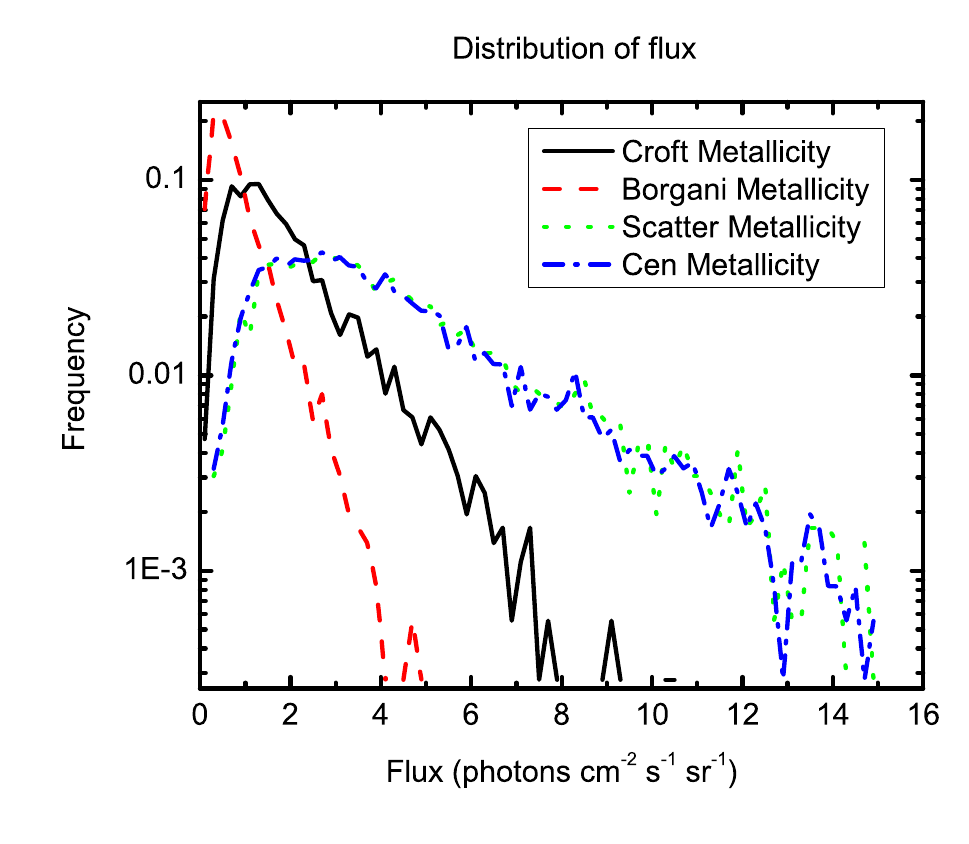}
\caption{Distribution function of surface brightness of the WHIM for 
simulations with FOV of 3' for the four metallicity models considered. 
Borgani metallicity is \emph{dashed red}, Croft metallicity is 
\emph{solid black}, Scatter metallicity is \emph{dotted green}, and 
Cen metallicity is \emph{dot-dashed blue}.}
\end{center}
\label{silicone_frequency_flux}
\end{figure}

We also investigated how the potential capability of a WHIM dedicated
future mission like EDGE \cite{Piro09} or 
Xenia \cite{Hartmann09}. 
The \emph{Wide Field Imager} (WFI) on EDGE (or Xenia - values in brackets) 
has a proposed field of view with diameter $1.5^\circ$ ($1.5^\circ$), 
an angular resolution of 15'' (10''), an effective area of 580 
(1000)~cm$^2$ at 1~keV, and an energy resolution of 70~eV at 
1~keV (70~eV at 0.5~keV). The \emph{Wide Field Spectrometer} (WFS) on 
EDGE (Xenia) has a proposed field of view of $0.7^\circ\times0.7^\circ$ 
($1^\circ\times1^\circ$), an angular resolution of 3.7' (2.5'), an 
effective area of 1163 (1300)~cm$^2$ at 600~eV, and an energy 
resolution of 3 (1)~eV at 0.5~keV. We simulated an experiment similar 
to the WFS, with roughly same effective area (1000~cm$^2$) and angular 
resolution ($3'\times3'$ pixels), values close to the EDGE/Xenia goal and 
consistent with our previous work. 
Figure \ref{silicone_frequency_flux} shows the 
frequency of flux depending on the metallicity model. The frequency using 
the Borgani model is peaked at a much smaller value than the other 
models, and seldom finds pixels with more than 6\phot. The Scatter 
and Cen models behave almost identically. They have a broad 
distribution, centered at around 5\phot, and a long tail going over 
the 20\phot. The Croft model is somehow in between, with a peak at 
approximately 2\phot and a tail going up to 15\phot. The capability 
of the mission to detect and study the WHIM will strongly depend 
on the correct metal abundance. 
In the case of our simulated experiment, with an exposure time 
of 1~Msec, assuming a total galactic foreground plus extragalactic 
background of $\sim29.3$\phot in the $380-950$~eV band (obtained 
using the three-component model by McCammon et al. 2002) and an 
instrumental noise of $\sim0.5$\phot  (in the same energy band), the 
threshold for WHIM detection is $\sim0.6$\phot. The 
probability of finding objects above such threshold is $\sim50\%$ 
for the Borgani metallicity and $99\%$ for the Scatter metallicity. 
The chance of finding objects with surface brightness greater than 
$20\%$ of the DXB, equivalent to $\sim6$\phot, is 20\% for the Scatter 
and Cen models, and negligible for the Borgani model. 

\begin{figure}
\begin{center}
\includegraphics{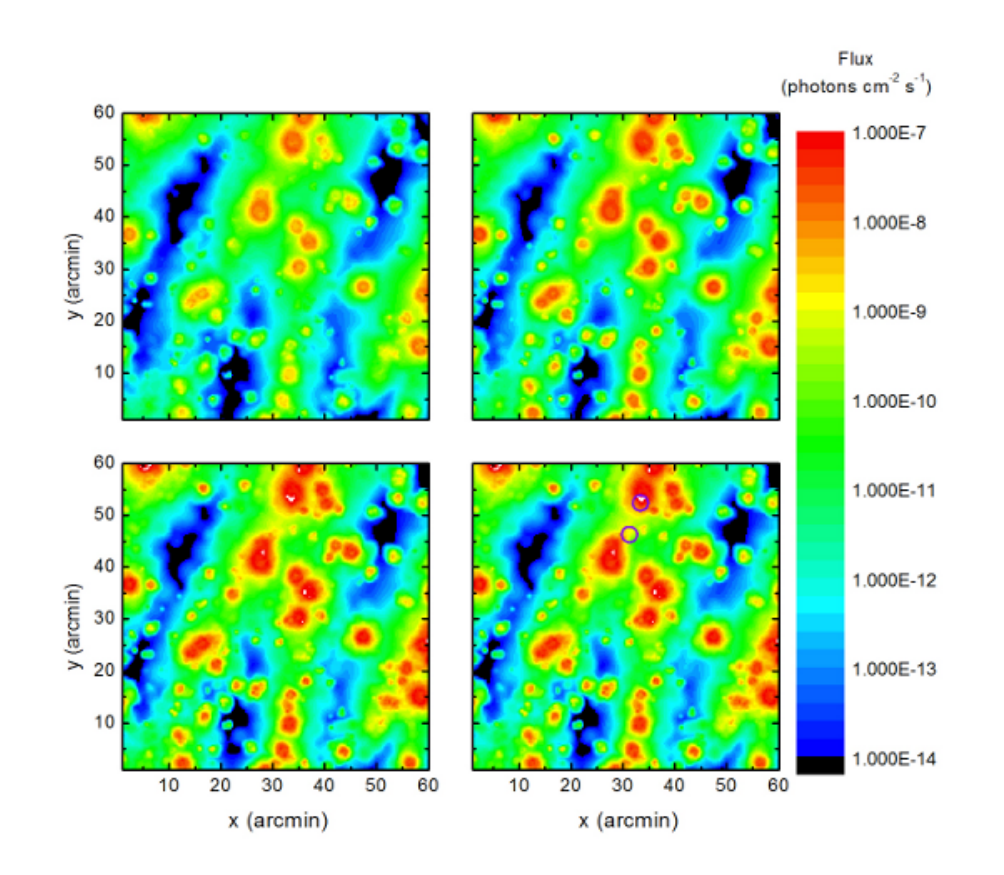}
\caption{Simulated WHIM maps at redshift $z=0.13-0.20$ with the 
Borgani metallicity model (\emph{top left}), the Croft model 
(\emph{top right}), the Scatter model (\emph{bottom left}), and the 
Cen model (\emph{bottom right}). The black circles in the bottom 
right image are centered on a dark and a bright region used to 
extract the spectra in Fig. \ref{compare-spectra}. The maps have a 
FOV of $1^\circ\times1^\circ$ and $256\times256$ pixels (resolution 
of 14'') in the energy range $380 - 950$~eV.}
\end{center}
\label{map-4models}
\end{figure}

Fig.~\ref{map-4models} shows images of the WHIM at redshift $z=0.13-0.20$.  
The ring-like shape of the bright objects is 
just an artifact due to the fact that X-rays coming from particles 
classified as hot or dense gas, which are present inside these 
objects, are not included in these images. The figures represent 
images of the same part of the sky with X-rays simulated using 
respectively the Borgani (top left), Croft (top right) Scatter (bottom 
left), and Cen (bottom right) models.
Notice that the amplitude of surface brightness of the four maps 
changes by almost an order of magnitude going from the Borgani to 
the Cen metallicity model. The Croft metallicity is proportional 
to density, therefore brightness is a direct function of density 
squared. For the Scatter and Cen Models, the ratio between the mean 
value of metallicity at low and high density is smaller, and therefore 
the fainter objects are more visible. There is also the scatter 
effect that comes into play changing the brightness between 
regions with the same density. The Borgani model is strongly affected 
by the poor metal diffusion that leaves many particles depleted in 
metals, therefore there is limited contribution from lines, and the 
emission is characterized almost only by the continuum due to 
bremsstrhalung. 

\begin{figure}
\begin{center}
\includegraphics[width=8cm]{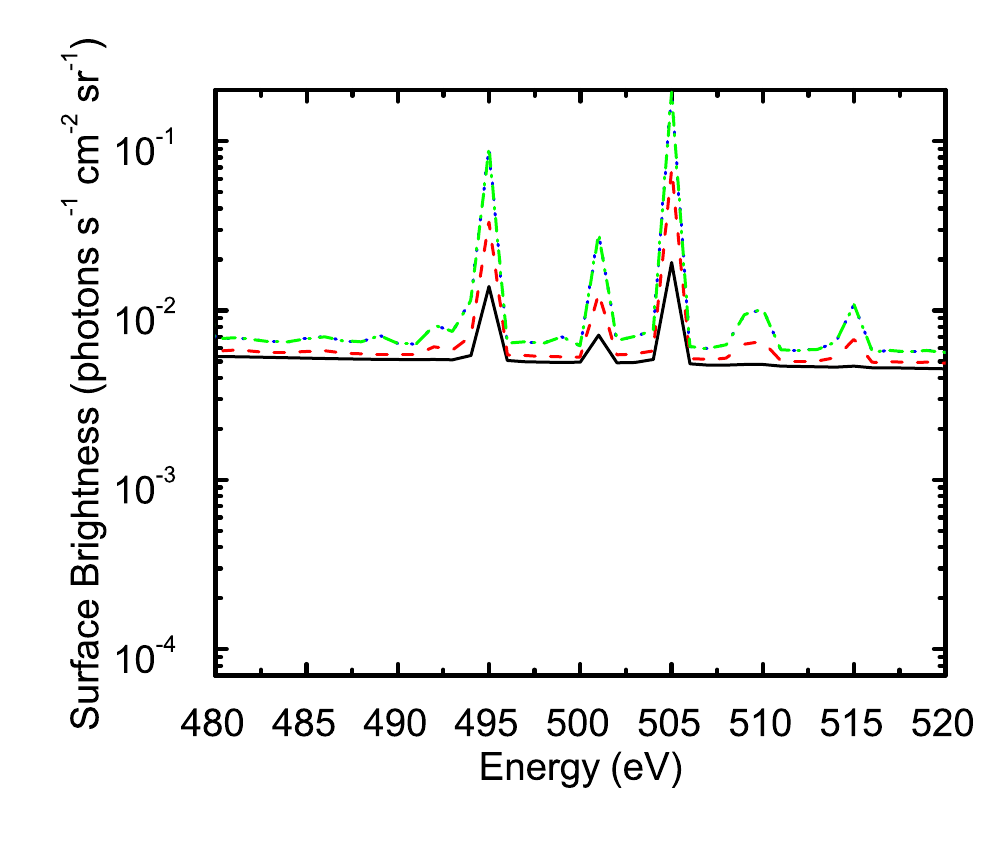}
\includegraphics[width=8cm]{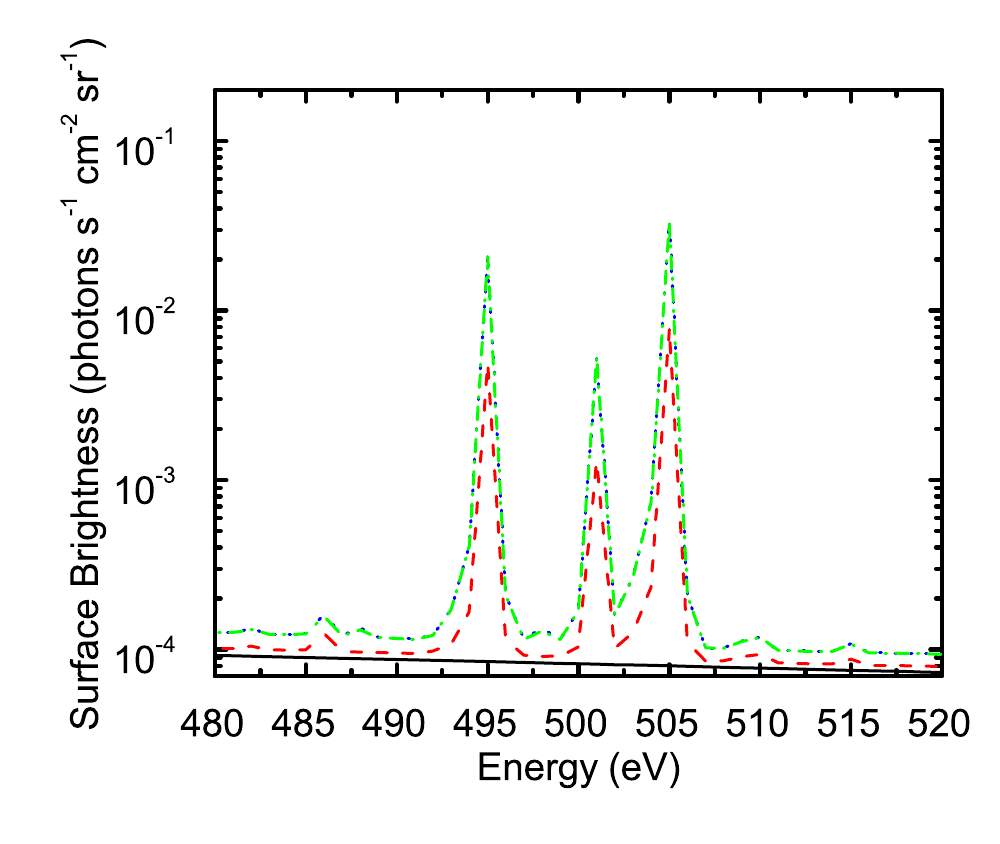}
\caption{Zoom in the $480-520$~eV of the simulated emission spectra from the selected 
dark and bright WHIM regions in the bottom right map of Fig. \ref{map-4models}. The
lines shown correspond to redhisfted O\elem{VII} triplet. 
The spectra are relative to simulations with the Borgani metallicity model 
(\emph{solid black line}), the Croft model (\emph{dashed red line}), the 
Scatter model (\emph{dotted green line}), and the Cen model 
(\emph{dot-dashed blue line}).}
\end{center}
\label{compare-spectra}
\end{figure}

In Fig.~\ref{compare-spectra} we show the energy spectra in the $480-520$~eV 
band extracted from two different regions of Fig.~\ref{map-4models} 
where the WHIM is particularly bright. The lines correspond to the O\elem{VII} 
triplet coming from two emitters at redshift $z\sim0.15$.  
Looking at the spectra we see how the metallicity model influences the  
intensity of the lines. The Scatter and Cen models have the highest 
metallicity and they predict the strongest emission line for 
an O\elem{VII} emitter. The lines simulated with the two models have 
almost the same intensity, with a little difference due to the 
correction to metallicity that the Cen model applies to the Scatter one. 
Since in this case the redshift is small and the overdensity 
is relatively high, the correction to metallicity is small. 
>From Fig.~\ref{metallicity-vs-density} we see that for the Croft model 
the metallicity is $3-5$ times lower than for the Scatter model, and 
the difference with the Cen model at higher redshift is smaller. The same 
ratio holds between the intensity of the lines in the spectra extracted and 
emission follows the lead of metallicity quite well. 
For the Borgani model things are different, as expected. The emitter in the 
left panel shows no sign of emission lines, even if simulations with the other 
models show an emitter at that position. 
Again, this is due to the poor metallicity distribution, this is a region 
where the particles of the simulation had no contact with metal rich particles 
and therefore have no metals at all, the emission is only in the continuum. 
The spectrum in the right panel shows the presence of O\elem{VII} lines 
also for the Borgani model, 
showing that this region has gone through metal enrichment, even if 
not as much as in the case of other metallicity models.

\begin{figure}
\begin{center}
\includegraphics{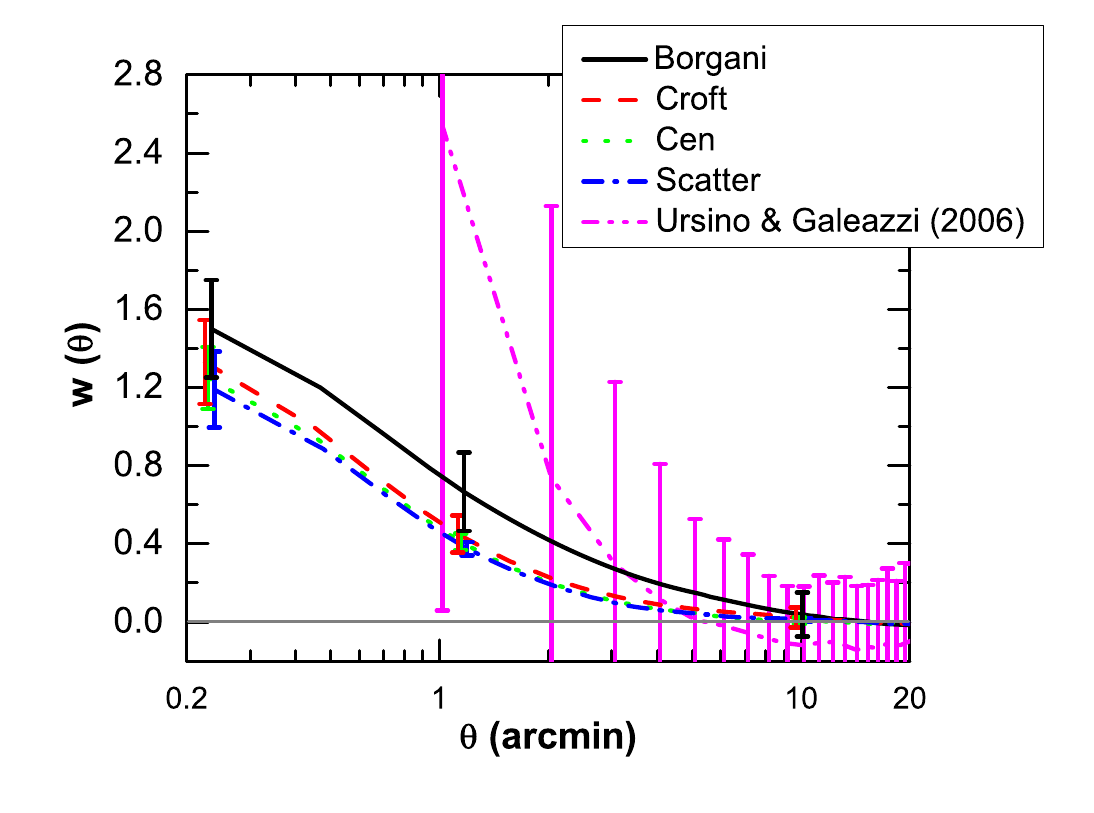}
\caption{Average angular autocorrelation function of the WHIM for the Borgani 
(\emph{solid black}), Croft(\emph{dashed red}), Cen (\emph{dottes green}), 
and Scatter (\emph{dot-dashed blue}) metallicity models compared with our previous 
work (\emph{double dot-dashed purple}). The error bars represent the cosmic variance 
for each set of maps.}
\end{center}
\label{AcFWHIM}
\end{figure}

Using maps along the full line of sight we can try to characterize the 
angular autocorrelation function \cite{Kuntz01} of the three phases, in order to extract the 
signal of the WHIM. So far experimental work has been done on the diffuse 
X-ray background \cite{Kuntz01,Soltan01,Giacconi01}, giving evidence of 
a signal at the order of less than 10~arcmin. In our previous work 
\cite{Ursino06} we used WHIM maps with rather broad angular resolution (1~arcmin) 
and found a characteristic angle comparable with the experimental values. 
With the current angular resolution of $14''$ we study the angular 
autocorrelation function (AcF) of the WHIM at even smaller scales, and how 
it evolves with metallicity. Fig.~\ref{AcFWHIM} shows the average AcFs for the 
Borgani, Croft, Scatter, and Cen metallicity models, with error bars 
representing the cosmic variance for each set of simulated maps. 
They are in good agreement 
with each other, giving a characteristic angle of $\sim10$~arcmin and well 
within the error bars (at this angle) of our previous work. The lower 
metallicity models hint to a somewhat slightly larger structure scale. 
The bad match with the older model has more than one reason aside from 
the very fact that we are dealing with different models. The older maps 
where smaller ($30'\times30'$ instead of $60'\times60'$) and the angular 
resolution was smaller ($1'\times 1'$, limited by the resolution of the
hydrodynamic simulation).
The small area gives rise to the much bigger variance and possibly to the 
fluctuation below 0 at $\theta>5$~arcmin. The worse angular resolution also 
explains (at least in part) the steeper slope at small angular scales.

\section{Conclusions}
\label{conclusion}
In our work we used the cosmological simulation by Borgani et al. 
(2004) to predict spectral and spatial properties of the WHIM. Since 
metallicity is the main source of uncertainties for this prediction, 
we used four  metallicity models, one self consistent with the Borgani 
simulation, and three analitical models, two of them based on statistics 
adopted from Cen \& Ostriker (1999a). The first model gives a lower 
limit to metallicity predictions, while the last two give a higher 
limit. For each model we generated a set of $1^\circ\times1^\circ$ 
maps up to redshift $z=2$. 

The tools we developed allowed us to characterize the WHIM emission, 
obtaining the following main results:
\begin{itemize}
	\item[(i)] The predicted X-ray emission from the WHIM depends strongly on the 
	metallicity model: the strongest emission comes from the more metallic 
	model, and it can be up to an order of magnitude stronger than from 
	the low metallicity models. The predicted emission, in particular 
	from high metallicity models, is in good agreement with observational 
	data. It accounts for a fraction between $2.5\%$ and $15\%$ of the total 
	DXB, and between $11\%$ and $66\%$ of the extragalactic emission at the 
	energy of O\elem{VIII} as measured from the XQC experiment \cite{McCammon02}. 
	It accounts for $8\%$ to $49\%$ of the extragalactic component 
	observed with ROSAT \cite{Kuntz01} and for $6.5\%$ to $33\%$ of what 
	measured with CHANDRA \cite{Hickox07}. The WHIM emission spans 
	from $10\%$ to $17\%$ of the total emission, in agreement with the 
	$12\%$ estimated with analysis of the AcF \cite{Galeazzi09}.
	\item[(ii)] The predicted surface brightness decreases with increasing redshift 
	of the emitting gas and most of the photons come from redshift $z<1.2$.
	\item[(iii)]The probability distribution function of emission along a line of sight
	depends on metallicity. For low metallicity the distribution is narrow and 
	peaked around its low average value, for high metallicity the distribution 
	is wide and there are higher chances to point at bright objects ($20\%$ 
	possibility to find an object brighter than 6\phot for the statistical models).
	\item[(iv)]A simple analysis of the maps with the AcF gives a 
	characteristic angular size of less than a few arcmin.
\end{itemize}

In summary, the prediction from our models are consistent with the 
constraints from observational data for all the metallicity distributions 
we assumed. Future mission with high energy resolution and good angular 
resolution will allow us to discriminate between the signals of the WHIM 
and of clusters of galaxies and possibly tracing the WHIM structure. 
Since the expected emission depends strongly on metallicity, measuring 
the flux from the WHIM will make possible to estimate the metal abundance 
of the WHIM.

\acknowledgments

This work has been supported in part by the University of Miami Small
Grant Program. The authors would like to thank Stefano Borgani for the
access to his hydrodynamic simulations, and to Lauro Moscardini and
Enzo Branchini for the useful discussion and suggestions.

\clearpage

\clearpage

\clearpage

\clearpage

\clearpage

\clearpage

\clearpage

\clearpage

\clearpage

\clearpage

\clearpage

\clearpage

\clearpage

\clearpage

\begin{deluxetable}{ccccc}
\tablewidth{0pt}
\tablecaption{Average WHIM metallicity, surface brightness (SF) in the
energy range 0.380-0.950~keV (in units of \phot), and predicted 
flux in the RASS $R4$ and $R5$ bands in units of 
10$^{-6}$~photons~s$^{-1}$~arcmin$^{-2}$ (the default RASS units). 
The metallicity is calculated at redshift 0.  
\label{surf-bright}}
\tablehead{
\colhead{Model} & \colhead{$\frac{Z}{Z_\odot}$} & \colhead{SF}
& \colhead{RASS R4 flux} & \colhead{RASS R5 flux}
}

\startdata
Borgani & 0.103 & $0.77\pm0.04$ & $1.2\pm 0.2$ & $1.3\pm 0.3$ \\
Croft & 0.043 & $1.84\pm0.08$ &  $3.4\pm 0.6$ & $3.1\pm 1.7$ \\
Scatter & 0.158 & $4.3\pm0.2$ &  $8.4\pm 1.5$ & $7.2\pm 1.7$ \\
Cen & 0.158 & $4.2\pm0.2$ &  $8.2\pm 1.6$ & $7.1\pm 1.7$ \\
\enddata	
\end{deluxetable}

\end{document}